\documentclass[letterpaper, 10 pt, conference]{ieeeconf}   
\IEEEoverridecommandlockouts                               
\usepackage{optidef}

\usepackage{amssymb,amsmath,color,subfigure}
\usepackage{graphicx}
\usepackage{xspace,stackrel,hyperref}
\usepackage{wrapfig}
\usepackage{mathtools}
\usepackage{tikz,pgfplots}
\usetikzlibrary{shapes,snakes}
\usepackage{etoolbox}
\usetikzlibrary{external} 
\usepackage{autonum}
\usepackage{dsfont}
\usepackage{etoolbox}
\usepackage{multirow}
\usepackage{cite}

\usepackage{algorithm, algpseudocode}
\usetikzlibrary{arrows.meta}
\usetikzlibrary{external}
\usepgfplotslibrary{fillbetween}
\usepgfplotslibrary{groupplots}
\usetikzlibrary{decorations.pathreplacing}
\usepackage[font=small,skip=2pt]{caption}
\usepackage{lipsum,adjustbox}

\definecolor{mycolor4}{RGB}{230,97,1}
\definecolor{mycolor2}{RGB}{178,171,210}
\definecolor{mycolor3}{RGB}{253,184,99}
\definecolor{mycolor1}{RGB}{94,60,153}

\makeatletter 
\pretocmd\@bibitem{\color{black}\csname keycolor#1\endcsname}{}{\fail}
\newcommand\citecolor[1]{\@namedef{keycolor#1}{\color{blue}}}
\makeatother

\DeclareMathAlphabet{\mathcal}{OMS}{cmsy}{m}{n}

\def\qed{\hfill \vrule height 7pt width 7pt depth 0pt\medskip}
\def\beq{\begin{equation}}
\def\eeq{\end{equation}}

\definecolor{mycolor1}{RGB}{230,97,1}
\definecolor{mycolor2}{RGB}{178,171,210}
\definecolor{mycolor3}{RGB}{253,184,99}
\definecolor{mycolor4}{RGB}{94,60,153}
\definecolor{mycolor5}{rgb}{0,0,0}

\newtheorem{assumption}{Assumption}

\newtheorem{theorem}{Theorem}
\newtheorem{proposition}{Proposition}

\newtheorem{definition}{Definition}

\newtheorem{remark}{Remark}

\newtoggle{full_version}
\toggletrue{full_version}
\title{\LARGE \bf
Moving Horizon Estimation for Simultaneous Localization and Mapping with Robust Estimation Error Bounds
}

\author{Jelena Trisovic, Alexandre Didier, Simon Muntwiler, Melanie N. Zeilinger% <-this % stops a space
\thanks{All authors are members of the Institute for Dynamic Systems and Control, ETH
Zurich, Zurich 8092, Switzerland. {\tt \{tjelena, adidier, simonmu,  mzeilinger\}@ethz.ch}. 
\newline Jelena Trisovic is supported by ETH AI center.}%
\iftoggle{full_version}{}{\thanks{An extended version containing all the proofs is available at \url{}}}%
}

\begin{document}

\maketitle
\thispagestyle{empty}
\pagestyle{empty}

\begin{abstract}
This paper presents a robust moving horizon estimation (MHE) approach with provable estimation error bounds for solving the simultaneous localization and mapping (SLAM) problem.  We derive sufficient conditions to guarantee robust stability in ego-state estimates and bounded errors in landmark position estimates, even under limited landmark visibility which directly affects overall system detectability. This is achieved by decoupling the MHE updates for the ego-state and landmark positions, enabling individual landmark updates only when the required detectability conditions are met. The decoupled MHE structure also allows for parallelization of landmark updates, improving computational efficiency. We discuss the key assumptions, including ego-state detectability and Lipschitz continuity of the landmark measurement model, with respect to typical SLAM sensor configurations, and introduce a streamlined method for the range measurement model. Simulation results validate the considered method, highlighting its efficacy and robustness to noise.

\end{abstract}

\section{Introduction}

Simultaneous localization and mapping (SLAM) refers to the fundamental task of enabling a robot to localize itself while concurrently constructing a map of an unknown environment using measurements of both robot and environment states.  Traditionally, SLAM is approached via filtering-based methods,  as reviewed in, e.g., \cite{bailey2006simultaneous,durrant2006simultaneous},  such as extended Kalman filters (EKFs) and particle filters (PFs), or optimization-based techniques. While EKFs are computationally efficient, they struggle with scalability and consistency, particularly with an increasing number of landmarks \cite{strasdat2012visual, julier2001counter}. Although PF-based methods are often more accurate, they are computationally intensive \cite{strasdat2012visual}. In contrast, optimization-based methods, also known as smoothing approaches, reframe SLAM as a maximum a posteriori (MAP) estimation problem that can be solved as sparse inference over factor graphs \cite{910572}, rendering them computationally less expensive.

Recent SLAM literature \cite{campos2021orb, sumikura2019openvslam, labbe2019rtab, 9793570} underscores the superiority of optimization-based methods and formulates SLAM as a constrained nonlinear least squares problem \cite{dellaert2017factor}. This problem can be solved using the measurements obtained over the robot's entire trajectory --- 
 an approach known in the control and estimation community as full information estimation (FIE). Alternatively, it can also be solved over selected measurements only, by considering ``key frames'' that are chosen as representative  \cite{campos2021orb}, or over a sliding window of consecutive, most recent measurements \cite{flayac2023distributed}. The latter technique is known in the estimation literature as moving horizon estimation (MHE).

%Simon's paper: [39]–[41]

MHE is commonly used for nonlinear estimation tasks involving noise and system constraints, providing robust stability guarantees when certain detectability conditions are met \cite{rawlings2017model}. In particular, choosing a sufficiently long horizon can ensure uniform robust stability of MHE under the general detectability condition  known as incremental input/output-to-state stability (i-IOSS), see, e.g.,\cite{schiller2023lyapunov,allan2021robust}. A particularly relevant direction in MHE research for SLAM applications is joint state and parameter estimation, of which SLAM is a special case, insofar as stationary landmarks are treated as a special case of system parameters. This general estimation problem is typically handled by augmenting the state to include both the system’s original states and unknown parameters \cite{zanon2013nonlinear, kuhl2011real, valluru2017development}. However, uniform detectability of this augmented state generally requires a persistent excitation (PE) condition \cite{schiller2023moving, sui2011moving}, which is particularly challenging in SLAM, due to the intermittent visibility of landmarks. Practical stability of MHE in the absence of PE has been recently established in \cite{muntwiler2023mhe}, while a switching mechanism depending on the level of PE was proposed in \cite{schiller2023nonlinear}.
However, these approaches are computationally complex for large numbers of features and verifying PE online is challenging in general.
Obtaining computationally efficient MHE approaches and simple conditions to verify PE online in the context of SLAM remains an open question, which we aim to address in this work.

This paper analyzes the estimation error properties in the context of MHE applied to SLAM, particularly with respect to the crucial issue of time-varying system detectability due to limited visibility of landmarks. This is achieved by harnessing the  theoretical guarantees regarding robustness of MHE for joint state and parameter estimation, as established in, e.g., \cite{schiller2023moving, muntwiler2023mhe}, and tailoring them specifically for challenges arising from SLAM. Consequently,  we consider a decoupled MHE scheme for SLAM similar to \cite{flayac2023distributed} and show robust stability of the ego-state estimates as well as robust error bounds on the landmark estimates. To the best of our knowledge, this is the first robust analysis for observers in the SLAM problem. In Section \ref{sec:1}, we outline the theoretical MHE foundations, followed by the general SLAM problem formulation in Section \ref{sec:2}. The decoupled MHE-SLAM scheme, detailed in Section \ref{sec:3}, relies on a detectability condition to ensure stable estimation of the ego-state at each time step, while the feature estimates are only updated if the available measurements are sufficiently informative. We further contextualize these conditions for different SLAM configurations in Section \ref{sec:4} and present a simplified scheme for the range measurement model, reducing the problem to a recursive least squares formulation. Section \ref{sec:6} provides simulation results that validate the decoupled MHE formulation.

\textit{Notation:} The sets of all (nonnegative) real numbers are denoted by ($\mathbb{R}_{\geq 0}$) $\mathbb{R}$, the set of integers in the interval $[a, b]$  by $\mathbb{I}_{[a, b]}$ and the set of integers greater or equal to $a$ by $\mathbb{I}_{\geq a}$. The bold symbol $\mathbf{u}$ refers to a sequence of the vector-valued variables $u_i \in \mathbb{R}^m$ for $i \in \mathbb{I}_{\geq 0}$, $ \mathbf{u}=\left\{u_0, u_1, \ldots\right\}$. Euclidean norm of vector $x \in$ $\mathbb{R}^n$ is denoted by $\|x\|$ and $\|x\|_P^2=x^{\top} P x$ for a positive definite matrix $P=P^{\top}$. The maximum generalized eigenvalue of positive definite matrices $P, Q$ is represented by $\lambda_{\max }(P, Q)$, i.e., the largest scalar $\lambda$ satisfying $\det (P - \lambda Q)=0$. The relation $P \succ 0$ $(P \succeq 0)$ denotes a positive definite (positive semi-definite) matrix.  The floor of $x \in \mathbb{R}$ is denoted by $ \lfloor x \rfloor$.The identity matrix is denoted as $I_n \in \mathbb{R}^{n \times n}$ and the zero square matrix as $O_n \in \mathbb{R}^{n \times n}$.

\section{Preliminaries}
\label{sec:1}
In this section, we review the MHE formulation and the corresponding theoretical guarantees from \cite{schiller2023lyapunov}.
For this purpose, we consider a discrete-time system of the form
\begin{equation}
\begin{aligned}
x_{k+1} & = f(x_{k}, u_k, w_k), \\
y_{k} & = h(x_{k}, u_k, w_k),
\end{aligned} \label{eq:gen_sys}
\end{equation}
where $x_k \in \mathbb{X}$, $y_k \in \mathbb{Y}$, $w_k \in \mathbb{W}$, $u_k \in \mathbb{U}$ are system state, measurements, process and measurement noise, and input at time $k$, respectively. The sets $\mathbb{X} \subseteq \mathbb{R}^n, \mathbb{Y} \subseteq \mathbb{R}^p, \mathbb{W} \subseteq \mathbb{R}^q, \mathbb{U} \subseteq \mathbb{R}^r$ are known closed sets that can arise from the physical nature of the system or some prior knowledge about the system trajectories. The functions $f: \mathbb{X} \times \mathbb{U} \times\mathbb{W} \rightarrow \mathbb{X}$  and $h: \mathbb{X} \times \mathbb{U} \times\mathbb{W} \rightarrow \mathbb{Y}$ are known nonlinear continuous functions that describe the system dynamics and measurement functions, respectively. Furthermore, we define $\mathbb{Z} \coloneqq  \left\{ (x,u,w) \in \mathbb{X} \times \mathbb{U} \times \mathbb{W}: f(x,u,w) \in \mathbb{X} \right\}$.
%, denoted as $ \mathbf{Y}_k = \{y_0, y_1, \dots, y_{k}\}$

Given the sequence of system measurements up to time $k$ and an initial estimate of system state, $\hat{x}_0$, the objective of an MHE scheme is to provide an estimate $\hat{x}_k$ of the current system state $x_k$ in a recursive manner, by only optimizing over a fixed window of measurements. For the underlying nonlinear estimation problem, we consider the following definition of detectability, as it is a sufficient and necessary condition for the existence of an observer with desired stability properties, as  described in Definition \ref{def2} \cite{allan2021nonlinear}.
\begin{definition} \label{def1}
    (Definition 2.9 from \cite{allan2021nonlinear}) A system described by \eqref{eq:gen_sys} is detectable if it admits an i-IOSS Lyapunov function $V: \mathbb{X} \times \mathbb{X} \rightarrow \mathbb{R}_{\geq 0}$, that is, there exist $\underline{U}, \overline{U} \succ 0$, $Q, R \succeq 0$ and  $\eta \in[0,1)$ such that
$$
\begin{aligned}
\|x-\tilde{x}\|_{\underline{U}}^2 &\leq V(x, \tilde{x}) \leq\|x-\tilde{x}\|_{\overline{U}}^2 \\
 V(f(x, u, w), f(\tilde{x}, u, \tilde{w})) &\leq \eta V(x, \tilde{x})+\|w-\tilde{w}\|_{Q}^2 \\
& +\|h(x, u, w)-h(\tilde{x}, u, \tilde{w})\|_{R}^2
\end{aligned}
$$
for all $(x, u, w),(\tilde{x}, u, \tilde{w}) \in \mathbb{Z}$.
\end{definition} 

An MHE scheme that estimates states of a detectable system \eqref{eq:gen_sys} (according to Definition \ref{def1}) was proposed in, e.g., \cite{schiller2023lyapunov}. At time $k$, it considers past input and output data over a window of length $M_k=\min\{k,M\}$, with $M \in \mathbb{I}_{\geq 1}$. It optimizes over the initial state at the start of the window (i.e., $\hat{x}_{k-M_k\mid k}$) and the sequence of process and measurement noise $\hat{w}_{\cdot \mid k}$ by minimizing the following cost function
\begin{align}
& J_k \left(\hat{x}_{k-M_k\mid k}, \hat{w}_{\cdot \mid k}, \hat{y}_{\cdot \mid k}\right) =   2\eta^{M_k} \left\|\hat{x}_{k-M_k\mid k}-\hat{x}_{k-M_k}\right\|_{\overline{U}}^2
 \\ & +\sum_{j=1}^{M_k}\eta^{j-1} \left(2\left\|\hat{w}_{k-j \mid k}\right\|_{Q}^2 + \left\|\hat{y}_{k -j \mid k}-y_{k-j}\right\|_{R}^2 \right),
 \end{align}
where $\eta, \overline{U}, Q$ and $R$ are such that the property from Definition \ref{def1} holds for some Lyapunov function, and notation $\hat{x}_{j \mid k}$ denotes the estimated state for time $j$ computed at time $k$. The estimated noise and outputs are denoted analogously by $\hat{w}_{j \mid k}$ and $\hat{y}_{j\mid k}$.
A state estimate is then obtained by solving the following nonlinear program (NLP) at each time step $k$ 
\begin{equation}
 \begin{alignedat}{2}
 \min _{\hat{x}_{k-M_k\mid k}, \hat{w}_{\cdot \mid k}} & J_k \left(\hat{x}_{k-M_k\mid k},  \hat{w}_{\cdot \mid k}, \hat{y}_{\cdot \mid k} \right) \\
 \text { s.t. } \hat{x}_{j+1 \mid k}&=f\left(\hat{x}_{j \mid k}, u_j, \hat{w}_{j \mid k}\right), && j \in \mathbb{I}_{\left[0, M_k-1\right]} \text {, } \\
  \hat{y}_{j \mid k} &= h\left(\hat{x}_{j \mid k}, u_j, \hat{w}_{j \mid k}\right), && j \in \mathbb{I}_{\left[0, M_k-1\right]}. \\
\end{alignedat}
\label{eq:MHEgeneral}
\end{equation}
The minimizer to \eqref{eq:MHEgeneral} is denoted as $\hat{x}^{*}_{k-M_k\mid k}$ and $\hat{w}^*_{\cdot \mid k}$, while the system state estimate at time step $k$ is given by $\hat{x}_{k}=\hat{x}^{*}_{k \mid k}$. In the case that sets  $ \mathbb{X}$ and $ \mathbb{W}$ are known and different from $\mathbb{R}^n$ and $\mathbb{R}^q$, additional constraints can be included in \eqref{eq:MHEgeneral}. In the following definition, we introduce the notion of robust global exponential stability, which serves to provide a robust bound on the estimation error.

\begin{definition} \label{def2} (Definition 2 from \cite{schiller2023lyapunov}) A state estimator for system \eqref{eq:gen_sys} is robustly globally exponentially stable (RGES) if there exist constants $\lambda_1, \lambda_2 \in [0,1)$ and $C_1,C_2 > 0$ such that the resulting state estimate $\hat{x}_k$ at time $k$ satisfies
\begin{equation}
\begin{aligned}
\left\|x_k-\hat{x}_k\right\| \leq & \max \biggl\{ C_1\lambda_1^k\left\|x_0 - \hat{x}_0\right\| \biggr., \\ &\left.\max _{j \in \mathbb{I}_{[0, k-1]}}\left\{ C_2\lambda_2^{k-j-1}\left\|w_{j}\right\|\right\}\right\}
\end{aligned}
\label{eq:def1}
\end{equation}
for all $k \in \mathbb{I}_{\geq 0}$, all initial conditions $x_0, \hat{x}_0 \in \mathbb{X}$, and every trajectory $(x_k, u_k, w_k, y_k)_{k=0}^{\infty}$ satisfying \eqref{eq:gen_sys}.
\end{definition}

Therefore, RGES implies that the estimation error decreases exponentially with time, down to an error that depends on the encountered process and measurement noise. It has been shown in \cite{schiller2023lyapunov} that RGES of the MHE scheme \eqref{eq:MHEgeneral} for a detectable system can be established as follows.
\begin{theorem} \label{th1}
    (Corollary 1 from \cite{schiller2023lyapunov}) Let system \eqref{eq:gen_sys} be detectable, i.e., admit an i-IOSS Lyapunov function according to Definition \ref{def1}. Then the MHE scheme \eqref{eq:MHEgeneral}  with horizon $M$ chosen such that $4\eta^M\lambda_{\max}(\underline{U}, \overline{U}) < 1$ is RGES according to Definition \ref{def2}.
\end{theorem}

\section{Problem Formulation}
\label{sec:2}

In this work, we consider static environments, as is standard in most SLAM literature. We distinguish between the ego-state $x^s \in \mathbb{X}^s \subseteq \mathbb{R}^{n_s}$ and the state of the environment $x^e$ it operates in, where the environment is described by a set of $L$ landmarks and their positions $x^{e,l} \in \mathbb{X}^{e,l} \subseteq \mathbb{R}^{n_e} $ for $l \in \mathbb{I}_{[1,L]}$, i.e., ${{x}^e} = ({x^{e,1\top}}, x^{e,2\top}, \dots, x^{e,L\top})^{\top}$. A distinction is made between measurements $y^e \in \mathbb{R}^{d}$, which rely on the environment (such as lidar, camera, etc.) and $y^s$, which do not (GPS measurements, measurements of distance from known anchor points, orientation measurements, etc.). The resulting nonlinear discrete-time system formulation is described by the following state space model  

\begin{subequations}\label{eq:slam_sys}
\begin{align}
    x^s_{k+1} & = f^s(x^s_{k}, u_k, v^s_k), \label{eq:slam_sys1} \\
    y^s_{k} & = h^s(x^s_{k}, \xi^s_k), \label{eq:slam_sys2} \\
    y^e_{k} & = A_k {H}^e(x^s_{k}, x^e, \xi^e_k), \label{eq:slam_sys3}
\end{align}
\end{subequations}
where $A_k \in \mathbb{R}^{dL \times dL}$ is a block diagonal matrix consisting of $L$ data association parameters $a_k^{e,l}$ on its block diagonal, with
%, indicating whether landmark $l$ is visible at time $k$:
\begin{equation}
    a_k^{e,l} = \begin{cases}
    I_d,& \text{if landmark } l \text{ is visible at time } k,\\
    O_d,              & \text{otherwise.}
\end{cases}
\end{equation}

Furthermore, the landmark measurement vector consists of stacked measurements of individual landmark positions: ${H}^e(x^s_{k}, x^e, \xi^e_k) = (h^{e}(x^s_{k}, x^{e,1}, \xi^{e,1}_k)^{\top}, \cdots, $ $ h^{e}(x^s_{k}, x^{e,L}, \xi^{e,L}_k)^{\top})^{\top}$, for  environment measurement noise $ \xi^e_k=(\xi^{e,1\top}_k,\cdots,\xi^{e,L\top}_k)^{\top}$, where $\xi^{e,l}_k \in {\mathbb{W}}^{e,l} \subseteq \mathbb{R}^{n_{me}}$. The measurement of a single landmark is therefore given by $y^{e,l}=a_k^{e,l}h^{e}(x^s_{k}, x^{e,l}, \xi^{e,l}_k)$ for $l \in \mathbb{I}_{[1,L]}$. The combined process and measurement noise vector of the system \eqref{eq:slam_sys1}-\eqref{eq:slam_sys2} is denoted as $w^s=(v^{s\top}, \xi^{s\top})^{\top}$, and analogously, the combined noise vector of the system \eqref{eq:slam_sys} as $w =  (v^{s\top},\xi^{s\top},\xi^{e\top})^{\top} $.

Note that we only consider the back-end of the SLAM pipeline and assume that the front-end correctly assigns measurements to the corresponding landmark. However, data association errors can be integrated into estimation schemes as shown in \cite{dellaert2017factor}. 

Given the sequence of system and environment measurements up to time $k \in \mathbb{I}_{\geq 0}$, as well as initial guesses of the ego-state $\hat{x}^s_0$ and landmark positions $\hat{x}^e_0$, the objective of the SLAM problem is to estimate the robot's trajectory $\mathbf{X^s} = \{x^s_0, x^s_1, \dots, x^s_{k}\}$ and landmark positions  ${{x}^e}$. This problem can be addressed using FIE to optimize over a pose graph, which is, however, not computationally tractable for large task horizons. Therefore, we formulate the SLAM inference problem within the MHE framework \eqref{eq:MHEgeneral}. We do so by considering the augmented system state $x=(x^{s\top}, x^{e\top})^{\top}$, which allows us to bring the system \eqref{eq:slam_sys} to the standard form of system \eqref{eq:gen_sys}. As a result, \eqref{eq:MHEgeneral} can be solved for $\hat{x}_{k - M_k \mid k}=(\hat{x}^{s\top}_{k-M_k \mid k}, \hat{x}^{e\top}_{k\mid k})^{\top}$ and $\hat{w}_{\cdot \mid k}=(\hat{v}^{s\top}_{\cdot \mid k}, \hat{\xi}^{s\top}_{\cdot \mid k}, \hat{\xi}^{e\top}_{\cdot \mid k})^{\top}$.

The resulting MHE problem can be seen as a special case of parametric estimation problems as considered in \cite{muntwiler2023mhe, schiller2023moving} since the landmark positions remain constant over time. Specifically, in the SLAM case the landmarks do not affect the system dynamics but rather provide additional measurements of interest.  As shown in \cite{schiller2023moving}, the detectability of the augmented state is equivalent to detectability of the system state (with respect to parameters and all measurements) and parameter PE. Hence, although the MHE scheme \eqref{eq:MHEgeneral}  provides a viable approach for addressing the SLAM problem, the resulting estimation error is RGES only when the trajectory of the robot is such that the landmarks are persistently excited. An alternative, decoupled MHE-based SLAM scheme was proposed in \cite{flayac2023distributed}. While its stability in the presence of noise was not rigorously analyzed, it was applied under the assumption that the trajectory of the system ensures PE of landmarks, rendering the scheme stable. Since this cannot be guaranteed in a general case as landmarks intermittently enter and exit the robot's field of view,  we analyze an MHE scheme inspired by \cite{flayac2023distributed}, and derive an upper bound on its error for a general trajectory that may not always ensure PE of landmarks. 

\section{Decoupled MHE for SLAM}
\label{sec:3}

In this section, we propose and analyze a decoupled MHE scheme for SLAM, where the ego-state is estimated solely based on system measurements, i.e., \eqref{eq:slam_sys2}, and the  positions of landmarks are estimated only based on environment measurements from \eqref{eq:slam_sys3} while keeping the ego-state fixed. The latter estimate is only performed when a condition ensuring sufficient informativeness of measurements is satisfied. We show that the scheme results in RGES estimates of the ego-state and bounded estimation error for landmark positions. Additionally, we show that the landmark estimation error bound decreases with the number of times the informativeness condition is satisfied. To ensure RGES of the MHE for state estimation, we require the following assumption.

\begin{assumption} \label{ass1}
 System \eqref{eq:slam_sys1}-\eqref{eq:slam_sys2} admits an i-IOSS Lyapunov function according to Definition \ref{def1}. 
\end{assumption}

Because the estimation of the landmark positions relies on the ego-state estimates and is thus also influenced by the state estimation error, we require the following assumption.

\begin{assumption} \label{ass2}
For any landmark $l$ with corresponding coordinates $x^{e,l}$, it holds that environment measurement function $h^e$ is Lipschitz continuous with respect to the ego-state, i.e., it holds that
\begin{equation}
    \left\|{h}^e\left({x}^s,{x}^{e,l},  {{\xi}}^{e,l}\right)-{{h}}^e\left(\tilde{x}^s, {{x}}^{e,l},  {{{\xi}}}^{e,l}\right)\right\| \leq C_L \left\| x^s - \tilde{x}^s \right\|
\end{equation}
for a constant $C_L \in \mathbb{R}_{\geq 0}$ and any $x^s, \tilde{x}^s \in \mathbb{X}^s, {x}^{e,l} \in \mathbb{X}^{e,l}$ and ${\xi}^{e,l} \in \mathbb{W}^{e,l}$.
\end{assumption}

We rewrite the landmark measurements as 
\begin{equation}
    {y}^{e,l}_{k} = a_k^{e,l}{h}^e\left(x_k^s, {x}^{e,l}, {{{\xi}}^{e,l}_{k}}\right) = a_k^{e,l}\overline{h}^e\left(\hat{x}^s_k, {x}^{e,l}, {{\overline{\xi}}^{e,l}_{k}}\right),
    \label{eq:rewrite_meas}
\end{equation}
where  ${\overline{\xi}}^{e,l}_{k}=({\xi}^{e,l\top}_{k}, \hat{e}^{s\top}_k)^{\top} \in \overline{\mathbb{W}}^{e,l} \subseteq \mathbb{R}^{n_s+n_{me}}$. By Lipschitz continuity of $h$ with respect to $x^s$ and the properties of vector norms, it holds that $\|{\overline{\xi}}^{e,l}_{k}\| \leq \overline{C}_1\|{\xi}_k^{e,l}\| + \overline{C}_2\|\hat{e}^{s}_k\| $ for some constants $\overline{C}_1, \overline{C}_2 \in \mathbb{R}_{\geq 0}$, with $\hat{e}^s_k\coloneqq x^s_k-\hat{x}^s_k$. In Section \ref{sec:4},  we verify this assumption for common sensor configurations.

In order to ensure a robust bound on the landmark estimation error, we update the landmark estimates only when informative measurements are available. To formulate this condition, we make use of the following definition for landmark detectability over a horizon of  length $M$, which is obtained by applying the  the exponential i-IOSS decrease condition from Definition \ref{def1}  $M$ times to a static system described by the position of a landmark and its measurement.

\begin{definition} \label{def3}
Landmark $l$ is detectable from measurements obtained over a horizon of length $M$ starting at time $k-M \in \mathbb{I}_{\geq 0}$, if  for some $\eta_{e,l} \in[0,1)$, $\underline{U}^{e,l}, \overline{U}^{e,l} \succ 0$, $Q^{e,l}, R^{e,l} \succeq 0$ and a given sequence of ego-state estimates  $\{\hat{x}^s_{k-M}, \dots, \hat{x}^s_{k-1}\}$,  the following property holds
\begin{equation}
\begin{aligned}
& \|{x}^{e,l}-\tilde{{x}}^{e,l}\|_{\underline{U}^{e,l}}^2  \leq \eta_{e,l}^{M}  \|{x}^{e,l}-\tilde{{x}}^{e,l}\|_{\overline{U}^{e,l}}^2 \\ & +\sum_{j=k-M}^{k-1} \eta_{e,l}^{k -j-1} \left(\left\|{\overline{\xi}}^{e,l}_{j} - \tilde{\overline{\xi}}^{e,l}_{j}\right\|_{Q^{e,l}}^2\right.  \\&  \left. + a_k^{e,l}\left\|\overline{h}^e\left(\hat{x}^s_{j},{x}^{e,l},  {\overline{\xi}}_j^{e,l}\right)-{\overline{h}}^e\left(\hat{x}^s_{j }, \tilde{{x}}^{e,l},  \tilde{{\overline{\xi}}}^{e,l}_{j}\right)\right\|_{R^{e,l}}^2 \right)
\end{aligned}
\label{eq:ass3}
\end{equation}
for any $x^{e,l}, \tilde{x}^{e,l} \in \mathbb{X}^{e,l}$ and ${{\overline{\xi}}}^{e,l}_{j},\tilde{{\overline{\xi}}}^{e,l}_{j} \in \overline{\mathbb{W}}^{e,l}$. We refer to such a horizon as \textit{informative} for landmark $l$ and denote it with $H_{k,M}^l$.
\end{definition}

 While evaluating the condition \eqref{eq:ass3} is in general a challenging problem, we discuss in Section \ref{sec:4} how it can be evaluated or ensured  for specific SLAM setups.

Given Assumptions \ref{ass1} and \ref{ass2}, we propose an MHE scheme described in Algorithm \ref{alg1}, which decouples estimation of the ego-state and landmark positions, allowing for parallelization of the latter. Additionally, the scheme enables the use of different time horizons for landmark and ego-state updates.

\begin{algorithm}[!b]
    \caption{\strut Decoupled SLAM MHE}
    \begin{algorithmic}[1]
    \State \textbf{Input}: $L, T, \hat{x}^e_{0}$,  $\hat{x}^s_{0}, N^s, N^{e,l}, \eta_s, \eta_{e,l}, \overline{U}^s, Q^s, R^s, \overline{U}^{e,l}$, 
    \\ $Q^{e,l}, R^{e,l} $, for $l \in \mathbb{I}_{[1,L]}$.  
    \State \textbf{Output: $\hat{x}^e_{T}$, $\hat{x}^s_T$}
    \For{$k = 1,2, \dots , T$}
    \State Obtain $y^s_k$ and $y^e_k$ from sensors
    \State Obtain $A_k$ from SLAM front-end
    \State $N_k^s \leftarrow \min \{k, N^s\}$ 
    \State  $\hat{x}^{s}_{k} $ $\leftarrow$ Solve \eqref{eq:sys_MHE} with $N_k^s, \eta_s, \hat{x}^{s}_{k - N_k^s} $
    \For{$l = 1,2, \dots , L$}  \algorithmiccomment{done in parallel}
    \If{\eqref{eq:ass3} holds for landmark $l$}
    \State $\hat{x}^{e,l}_{k} \leftarrow$ Solve \eqref{eq:Pkle} with $N^{e,l}, \eta_{e,l}, \hat{x}^{s}_{k}, \hat{x}^{e,l}_{k - N_k^s} $
    \Else
    \State $\hat{x}^{e,l}_{k} \leftarrow \hat{x}^{e,l}_{k-1}$
    \EndIf
    \EndFor
    \EndFor
    \end{algorithmic}
  \label{alg1}
\end{algorithm}

The cost which is minimized at every time step to obtain the ego-state estimate is given by
\begin{align}
& J_k^s\left(\hat{x}^s_{k-N_k^s \mid k}, \hat{w}^s_{\cdot \mid k}\right) =2\eta_s^{ N_k^s}\left\|\hat{x}^s_{k-N_k^s \mid k}-\hat{x}^s_{k-N_k^s}\right\|_{\overline{U}^s}^2 \\ & +\sum_{j=1}^{N_k^s} \eta_s^{j-1}\left(2\left\|\hat{w}^s_{k-j \mid k}\right\|_{Q^s}^2+\left\|\hat{y}^s_{k-j \mid k}-y^s_{k-j}\right\|_{R^s}^2 \right),
\end{align}
where  $\eta_{s} \in[0,1), \overline{U}^s, Q^{s}$ and $R^{s}$ correspond to $\eta \in[0,1)$, $\overline{U}, Q$ and $R$ in Assumption \ref{ass1}. The window length is set to $N^s_k = \min \{k, N^s\}$, where $N^s$ is chosen such that $4 \eta_s^{N^s} \lambda_{\max }\left(\overline{U}^s, \underline{U}^s\right) < 1$. To obtain the ego-state estimate at time $j$, computed at time $k$, denoted with  $\hat{x}^s_{j\mid k}$, as well as noise and output estimates $\hat{w}_{j\mid k}^s$ and $\hat{y}^s_{j\mid k}$, the following NLP is solved at every time step
\begin{equation}
    \begin{alignedat}{2}
& \min _{\hat{x}^s_{k-N_k^s \mid k}, \hat{w}^s_{\cdot \mid k}} J_k^s\left(\hat{x}^s_{k-N_k^s \mid k}, \hat{w}^s_{\cdot \mid k}\right) \\
 &\text { s.t. }  \hat{x}^s_{j+1 \mid k} =f^s\left(\hat{x}^s_{j \mid k}, u_j, \hat{w}^s_{j \mid k}\right), &&j \in \mathbb{I}_{\left[k-N_k^s, k-1\right]} \text {, } \\
& \hspace{1cm} \hat{y}^s_{j \mid k} =h^s\left(\hat{x}^s_{j \mid k}, u_j, \hat{w}^s_{j \mid k}\right), && j \in \mathbb{I}_{\left[k-N_k^s, k-1\right]}. % & \left(\hat{x}^s_{j \mid k}, u_j, {w}^s_{j \mid k}\right) \in \mathbb{Z}^s, j \in \mathbb{I}_{\left[k-N_k^s, k-1\right]}
\end{alignedat}\label{eq:sys_MHE}
\end{equation}
 The optimal state and noise estimates of the given NLP are denoted as $\hat{x}^{s*}_{\cdot \mid k}$ and $ \hat{w}^{s*}_{\cdot \mid k}$ respectively. The ego-state estimate at time $k$ is defined as $\hat{x}^s_k \coloneqq \hat{x}_{k \mid k}^{s*}$. 
 
% where notation    The objective of \eqref{eq:Pkle} is given by

To obtain the landmark position estimates  computed at time $k$, $\hat{x}^{e,l}_{k\mid k}$, and noise and output estimates at time $j$, estimated at time $k$, $\hat{\overline{\xi}}^{e,l}_{j\mid k}$ and $\hat{y}^{e,l}_{j\mid k}$, the following cost is minimized for each landmark $l$, possibly in parallel 
$$
\begin{aligned}
&  J_k^{e,l}\left(\hat{ {x}}_{k\mid k}^{e,l}, \hat{\overline{\xi}}^{e,l}_{\cdot \mid k},  \hat{{y}}^{e,l}_{\cdot \mid k}\right) =2\eta_{e,l}^{M}\left\|\hat{{x}}_{k\mid k}^{e,l}-{ \hat{x}}_{k-M}^{e,l}\right\|_{\overline{U}^{e,l}}^2 \\ & + \sum_{j=1}^{M} \eta_{e,l}^{j-1}\left(2\left\|\hat{\overline{\xi}}_{k-j \mid k}\right\|_{Q^{e,l}}^2+\left\|\hat{{y}}^{e,l}_{k-j \mid k}-{{y}}^{e,l}_{k-j}\right\|_{R^{e,l}}^2 \right).
\end{aligned}
$$

As described in Algorithm \ref{alg1}, and unlike \eqref{eq:sys_MHE}, the landmark position is updated only over informative horizons that satisfy \eqref{eq:ass3} for fixed weight matrices and decay rate, to avoid updating the landmark estimate with insufficiently informative data. In the case that the landmark detectability condition is not satisfied, the landmark  estimate is kept equal to the one from the previous step. The landmark estimate is obtained by solving the following NLP
\begin{equation}
    \begin{aligned}
& \min _{\hat{{x}}_{k \mid k}^{e,l}, \hat{\overline{\xi}}^{e,l}_{\cdot \mid k}} J_k^{e,l}\left(\hat{ {x}}_{k\mid k}^{e,l}, \hat{\overline{\xi}}^{e,l}_{\cdot \mid k},  \hat{{y}}^{e,l}_{\cdot \mid k}\right) \\
\text { s.t. } &  \hat{{y}}^{e,l}_{j \mid k}=a_k^{e,l}\overline{h}^e\left( \hat{x}^s_k,\hat{ x}_{k\mid k}^{e,l}, \hat{\overline{\xi}}^{e,l}_{j \mid k}\right), j \in \mathbb{I}_{\left[k-N^{e,l}_k, k-1\right]},
\end{aligned}\label{eq:Pkle}
\end{equation}
with horizon $M$ chosen such that $4 \eta_{e,l}^{M} \lambda_{\max }\left(\overline{U}^{e,l}, \underline{U}^{e,l}\right) < 1$ for every $l \in \mathbb{I}_{[1,L]}$. The optimal landmark position and noise of the given NLP are denoted as $\hat{x}^{e,l*}_{k \mid k}$ and $ \hat{\overline{\xi}}^{e,l*}_{\cdot \mid k}$ respectively. The landmark estimate at time $k$ is defined as $\hat{x}^{e,l}_k \coloneqq  \hat{x}_{k \mid k}^{e,l*}$. 

%horizon $N^{e,l} = M$ so that \eqref{eq:ass3} is satisfied. %\min \{k, M^{e,l}\}$.

In the following, we show RGES of ego-state estimates and bounded errors for landmark position estimates obtained by the proposed scheme, where the error bound for each landmark position directly depends on how often that landmark was detectable in the sense of Definition \ref{def3}. For every landmark $l \in \mathbb{I}_{[1,L]}$ and $j \in \mathbb{I}_{[1, m^{l}_k-1]}$, we denote $\tau^{l,k}_1 = \max\{{\tau}\leq k \mid H_{\tau, M}^l \text{ is informative }  \}$ and similarly, $\tau^{l,k}_{j+1} = \max\{{\tau} \leq \tau^{l,k}_{j}-M \mid H_{\tau, M}^l \text{ is informative }  \}$, where \textit{informativity index} $m^{l}_k \in  \mathbb{I}_{\geq 0}$ is such that $ \tau^{l,k}_{m_k^l}$ is the end of the earliset informative horizon for landmark $l$. As a result, the ends of disjunct informative horizons for landmark $l$, up to time $k$ correspond to $\tau^{l,k}_{j}$ for $j \in \mathbb{I}_{[1, m^{l}_k]}$.

\begin{proposition} \label{prop1}
    If Assumptions \ref{ass1} and \ref{ass2} are satisfied, the solution to the SLAM problem described by Algorithm \ref{alg1} and NLPs \eqref{eq:sys_MHE} and \eqref{eq:Pkle} has the following properties:
\begin{enumerate}
    \item the MHE approach \eqref{eq:sys_MHE} is RGES for system \eqref{eq:slam_sys1}-\eqref{eq:slam_sys2};
    \item for each landmark $l \in \mathbb{I}_{[1,L]}$, there exist constants $C_{1}^l, C_{2}^l,C_{3}^l,C_{4}^l > 0$ and $ \lambda_{1,l}, \lambda_{2,l}, \lambda_{3,l}, \lambda_{4,l} \in [0,1)$ such that landmark estimation error~$\hat{e}^{e,l}_k\coloneqq x^{e,l}-\hat{x}^{e,l}_k$ satisfies

\begin{equation}
\begin{aligned}
&\left\|\hat{e}^{e,l}_k\right\| \leq  \max \biggl\{ {C}_{1}^l{\lambda}_{1,l}^{m^{l}_kM}\left\|\hat{e}^{e,l}_0\right\|, m^{l}_k{C}_{2}^l{\lambda}_{2,l}^{m^{l}_kM}\left\|\hat{e}^{s}_0\right\| \biggr., \\
& \max _{t \in \mathbb{I}_{[0, m^{l}_kM-1]}} \left.\left\{  C_{3}^l\lambda_{3,l}^{m^{l}_kM-t-1}\left\|{\xi}^{e,l}_{s(t)^{l}_k}\right\| \right\}\right.,\\ &
 \max _{\substack{t \in \mathbb{I}_{[0, m^{l}_kM-1]},\\  q \in \mathbb{I}_{[0, s(t)^{l}_k-1]}}} \biggl.\left\{  C_{4}^l\lambda_{4,l}^{m^{l}_kM-t+q-1}\left\|{w}^{s}_{s(t)^{l}_k-q-1}\right\| \right\}\biggr\}
\end{aligned}
\label{eq:prop1}
\end{equation}
for all $k \in \mathbb{I}_{\geq 0}$ and $t$ denoting a counter over the time steps belonging to informative horizons $H_{\tau_j^{k,l}, M}^l$, $ j \in \mathbb{I}_{[1, m^{l}_k]}$. For some $t$, the corresponding time step is denoted with $s(t)^{l}_k = \tau_{m^{l}_k -\bigl \lfloor \frac{t}{M} \bigr \rfloor}^{k,l}-M+t - M\bigl \lfloor \frac{t}{M} \bigr \rfloor$\footnote{That means that $t=0$ corresponds to the first time step of the first informative horizon, i.e., $s(0)_k^l=\tau_{m_k^l}^k-M$, and that $t=m_kM-1$ corresponds to the last time step of the last informative horizon, i.e., $s(m_kM-1)_k^l=\tau_1^k-1$ .}. 
\end{enumerate}
\end{proposition}

\textit{Proof:} For the ego-state estimate, the estimation error is RGES due to Assumption \ref{ass1} and Theorem \ref{th1} applied to system \eqref{eq:slam_sys1}-\eqref{eq:slam_sys2}. 
 Starting from the landmark detectability condition \eqref{eq:ass3}, we relate the landmark estimation error to the optimal solution of the MHE cost in \eqref{eq:cost_connection}, similar to \cite{schiller2023lyapunov}. We then establish the decrease of the  error over each informative horizon, while its value does not change when  the horizon is not informative due to keeping the estimate constant in Line 13 of Algorithm \ref{alg1}, resulting in \eqref{eq:error_connection}. Due to Assumption \ref{ass2} and ego-state estimate being RGES, we obtain the error bound \eqref{eq:prop1}. Throughout this proof, we consider only a single landmark $l$ and hence drop $l$ from $m_k^l, \tau_j^{k,l}$ and all weight matrices' superscripts.
% Then, due to the rewritten measurement noise $\overline{\xi}^{e,l}$ directly depending on the state estimation error and original landmark measurement noise, we show the stated property of landmark estimation error. 
% $\delta_k$ denote the time that has passed since the most recent informative horizon ended, i.e., let $H^{\overline{k}}_{\overline{k}-N_{\overline{k}}^{e,l}}$ for

Since $\hat{e}^{e,l}_k=\hat{e}^{e,l}_{\tau_1^k}$ and the property \eqref{eq:ass3} is satisfied for the informative horizon $H_{\tau_1^k, M}^l$, the following holds for  the solution of the MHE problem \eqref{eq:Pkle} at time step ${\tau_1^k}$, $\left(\hat{x}_{{\tau_1^k}}^{e,l}, \{\hat{\overline{\xi}}^{e,l*}_{i \mid {\tau_1^k}}, \hat{{y}}^{e,l*}_{i\mid {\tau_1^k}}\}_{i={\tau_1^k}-M}^{{\tau_1^k}-1}\right)$ and the evolution of the true system state $\left({x}^{e,l}, \{{\overline{\xi}}^{e,l}_i, {{y}}^{e,l}_i\}_{i={\tau_1^k}-M}^{{\tau_1^k}}\right)$:
% (the first one is the solution of the MHE problem and therefore satisfies \eqref{eq:land_sys}, while the second one is the evolution of the true system state):
\begin{equation}
\begin{aligned}
& \|\hat{x}_{{\tau_1^k}}^{e,l} - {x}^{e,l}\|_{\underline{U}^{e}}^2  \leq \eta_{e,l}^{M}  \|\hat{x}_{{\tau_1^k}}^{e,l} - {x}^{e,l}\|_{\overline{U}^{e}}^2 \\ & {+}\!\!\sum_{j={\tau_1^k}-M}^{{\tau_1^k}-1} \eta_{e,l}^{{\tau_1^k}-j-1}\left(\left\|\hat{\overline{\xi}}^{e,l*}_{{j \mid {\tau_1^k}}} - {\overline{\xi}}^{e,l}_{j}\right\|_{Q^{e}}^2 + \left\| \hat{{y}}^{e,l*}_{j \mid {\tau_1^k}}-{{y}}^{e,l}_j\right\|_{R^{e}}^2\right).
\end{aligned}
% \label{eq:}
\end{equation}
Applying the Cauchy-Schwartz and Young's inequality, we obtain 
$$
\left\|\hat{\overline{\xi}}^{e,l*}_{{j \mid {\tau_1^k}}} - {\overline{\xi}}^{e,l}_{j}\right\|_{Q^{e}}^2 \leq 2\left\|\hat{\overline{\xi}}^{e,l*}_{{j \mid {\tau_1^k}}}\right\|_{Q^{e}}^2 + 2\left\|{\overline{\xi}}^{e,l}_{j}\right\|_{Q^{e}}^2
$$
and therefore it holds that:
\begin{equation}
\begin{aligned}
& \|\hat{x}_{{\tau_1^k}}^{e,l} - {x}^{e,l}\|_{\underline{U}^{e}}^2  \leq \eta_{e,l}^{M}  \|\hat{x}_{{\tau_1^k}}^{e,l} - {x}^{e,l}\|_{\overline{U}^{e}}^2  
\\& +2\sum_{j={\tau_1^k}-M}^{{\tau_1^k}-1} \eta_{e,l}^{{\tau_1^k}-j-1}\left(\left\|\hat{\overline{\xi}}^{e,l*}_{{j \mid {\tau_1^k}}} \right\|_{Q^{e}}^2 + \left\|{\overline{\xi}}^{e,l}_{j}\right\|_{Q^{e}}^2\right) 
%\\& +2\sum_{j=0}^{M-1} \eta_{e,l}^{M-j}
\\& +\sum_{j={\tau_1^k}-M}^{{\tau_1^k}-1} \eta_{e,l}^{{\tau_1^k}-j-1} \left\| \hat{{y}}^{e,l*}_{j \mid {\tau_1^k}}-{{y}}^{e,l}_j\right\|_{R^{e}}^2 .
\end{aligned}
\label{eq:8}
\end{equation}
Due to:
$$
\begin{aligned}
   & \|\hat{x}_{{\tau_1^k}}^{e,l} - {x}^{e,l}\|_{\overline{U}^{e}}^2 = \|\hat{x}_{{\tau_1^k}}^{e,l} - \hat{x}_{{\tau_1^k}-M}^{e,l} + \hat{x}_{{\tau_1^k}-M}^{e,l} -{x}^{e,l}\|_{\overline{U}^{e}}^2 \\
   & \leq 2\|\hat{x}_{{\tau_1^k}}^{e,l} - \hat{x}_{{\tau_1^k}-M}^{e,l}\|_{\overline{U}^{e}}^2 + 2\|\hat{x}_{{\tau_1^k}-M}^{e,l} -{x}^{e,l}\|_{\overline{U}^{e}}^2, \\
\end{aligned}
$$
 expression \eqref{eq:8} becomes (as $\hat{x}_k^{e,l}=\hat{x}_{k \mid k}^{e,l*}$ for any $k \in \mathbb{I}_{\geq 0}$)
\begin{equation}
\begin{aligned}
& \|\hat{x}_{{\tau_1^k} }^{e,l} - {x}^{e,l}\|_{\underline{U}^{e}}^2  \leq 2\eta_{e,l}^{M} \|\hat{x}_{{\tau_1^k}-M}^{e,l} -{x}^{e,l}\|_{\overline{U}^{e}}^2  
\\&{+}{2}\!\!\!\!{\sum_{j={\tau_1^k}-M}^{{\tau_1^k}-1}}\!\!\eta_{e,l}^{{\tau_1^k}{-}j{-}1}\left\|{\overline{\xi}}^{e,l}_{j}\right\|_{Q^{e}}^2{+}J_{{\tau_1^k}}^{e,l}\left(\hat{ {x}}_{{\tau_1^k} \mid {\tau_1^k}}^{e,l*}, \hat{x}^s_{{\tau_1^k}}, \hat{\overline{\xi}}^{e,l*}_{\cdot \mid {\tau_1^k}},  {\hat{y}}^{e,l*}_{\cdot \mid {\tau_1^k}}\right).
\label{eq:cost_connection}
\end{aligned}
% \label{eq:}
\end{equation}
Because of the feasibility of the true system trajectory in the MHE optimization problem, it holds that
$$J_{{\tau_1^k}}^{e,l}\left(\hat{ {x}}_{{\tau_1^k}\mid {\tau_1^k}}^{e,l*}, \hat{x}^s_{{\tau_1^k}}, \hat{\overline{\xi}}^{e,l*}_{\cdot \mid {\tau_1^k}},  \hat{{y}}^{e,l*}_{\cdot \mid {\tau_1^k}}\right){\leq} J_{{\tau_1^k}}^{e,l}\left({ {x}}^{e,l}, \hat{x}^s_{{\tau_1^k}}, {\overline{\xi}}^{e,l}_{\cdot \mid {\tau_1^k}},  {{y}}^{e,l}_{\cdot \mid {\tau_1^k}}\right)$$
which further implies (since $\hat{e}_{{\tau_1^k}}^{e,l} =\hat{x}_{{\tau_1^k}}^{e,l} - {x}^{e,l}$):
\begin{equation}
\begin{aligned}
& \|\hat{e}_{{\tau_1^k}}^{e,l}\|_{\underline{U}^{e}}^2  \leq 4\eta_{e,l}^{M} \|\hat{e}_{{\tau_1^k}-M}^{e,l}\|_{\overline{U}^{e}}^2  + 4\sum_{j={\tau_1^k}-M}^{{\tau_1^k}-1} \eta_{e,l}^{{\tau_1^k}{-}j{-}1}\left\|{\overline{\xi}}^{e,l}_{j}\right\|_{Q^{e}}^2
\\& \leq 4\overline{\lambda}\eta_{e,l}^{M}\|\hat{e}_{{\tau_1^k}-M}^{e,l}\|_{\underline{U}^{e}}^2 + 4\sum_{j={\tau_1^k}-M}^{{\tau_1^k}-1} \eta_{e,l}^{{\tau_1^k}{-}j{-}1}\left\|{\overline{\xi}}^{e,l}_{j}\right\|_{Q^{e}}^2,
\end{aligned}
% \label{eq:}
\end{equation}
where $\overline{\lambda} \coloneqq \lambda_{\mathrm{max}}({\overline{U}^{e}}, {\underline{U}^{e}})$. Due to the choice of horizon length $M$ in \eqref{eq:Pkle}, $\rho_{l} \in [0,1)$ exists so that $\rho_{l}^{M} \coloneqq  4\eta_{e,l}^{M}\overline{\lambda} < 1$. From here, we obtain the upper bound of the estimation error as follows:
\begin{equation}
\begin{aligned}
& \|\hat{e}_{{\tau_1^k}}^{e,l}\|_{\underline{U}^{e}}^2  \leq \rho_{l}^{M} \|\hat{e}_{{\tau_1^k}-M}^{e,l}\|_{\underline{U}^{e}}^2 +4\sum_{j={\tau_1^k}-M}^{{\tau_1^k}-1} \eta_{e,l}^{{\tau_1^k}-j-1}\left\|{\overline{\xi}}^{e,l}_{j}\right\|_{Q^{e}}^2
\end{aligned}
% \label{eq:} 
\end{equation}
% Let $\overline{k}=m_lM+p_lM+q_l$ for some $m_l, p_l,q_l \in \mathbb{N}$, where $m_l$ is the number of informative horizons and $q \in \mathbb{I}_{[0,M-1]}$.  

Similarly, since no update is performed at times that do not correspond to 
 $\tau^{k}_{j}$ for $j \in \mathbb{I}_{[1, m_k]}$, the error remains constant until  $\tau^{k}_{2}$, so we have  $\hat{e}^{e,l}_{\tau_{1}^k-M}=\hat{e}^{e,l}_{\tau_2^k}$. Therefore, we can repeat the same process and obtain the bound on $\|\hat{e}_{\tau_{1}^k-M}^{e,l}\|_{\underline{U}^{e}}^2$ as
\begin{equation}
\begin{aligned}
& \|\hat{e}_{\tau_{2}^k}^{e,l}\|_{\underline{U}^{e}}^2  \leq \rho_{l}^{M} \|\hat{e}_{\tau_{2}^k-M}^{e,l}\|_{\underline{U}^{e}}^2 +4\sum_{j=\tau_{2}^k-M}^{\tau_{2}^k-1} \eta_{e,l}^{\tau_{2}^k-j-1}\left\|{\overline{\xi}}^{e,l}_{j}\right\|_{Q^{e}}^2.
\end{aligned}
% \label{eq:} 
\end{equation}

Therefore, since the first informative horizon started at $\tau_{m_k}-M$, it holds that $\hat{e}_{\tau_{m_k}-M}^{e,l}=\hat{e}_{0}^{e,l}$, leading to
\begin{equation}
\begin{aligned}
& \|\hat{e}_{k}^{e,l}\|_{\underline{U}^{e}}^2  \leq \rho_{l}^{m_{k}M} \|\hat{e}_{0}^{e,l}\|_{\underline{U}^{e}}^2  
\\& +4\sum_{j=1}^{m_{k}}\rho_l^{jM} \sum_{r=\tau_j^k-M}^{\tau_j^k -1} \eta_{e,l}^{\tau_j^k-r-1}\left\|{\overline{\xi}}^{e,l}_{r}\right\|_{Q^{e}}^2.
\end{aligned}
\label{eq:error_connection} 
\end{equation}

From Assumption \ref{ass2}, $\|{\overline{\xi}}^{e,l}_{k}\|^2 \leq \overline{C}_1\|{\xi}_k^{e,l}\|^2 + \overline{C}_2\|\hat{e}^{s}_k\|^2$, where $\|\hat{e}^{s}_k\|^2$ is  upper bounded, due to MHE for state estimation from \eqref{eq:sys_MHE} being RGES. After plugging this upper bound into \eqref{eq:error_connection} and maximizing the sum  similarly to the proof of Corollary 1 in \cite{schiller2023lyapunov}, we obtain \eqref{eq:prop1}, where the decrease is given with respect to the informativity index $m_k$, as the error is constant if the landmark is not detectable. $\hfill \qed$

It therefore holds that the landmark estimation error is robustly stable, where the effect of initial ego-state and landmark estimation errors decreases exponentially with respect to informativity index $m_k^l$. The influence of past process and measurement noise on the error bound is discounted at the rate that also depends on the informativity index.
\begin{remark}
     Suppose that, for each landmark $l$, there exists a constant $T^{(l)} \in \mathbb{I}_{\geq 1}$ such that for every time $t \in \mathbb{I}_{\geq 1}$, there is a time $t_0 \in \mathbb{I}_{[t, t+(T^{(l)}-1)M]}$ for which $H_{t_0+M, M}^l$,  is informative, i.e., each landmark is periodically detectable over a horizon of length $M$. The informativity index is then proportional to time $k$, i.e., $m_k^l = \bigl \lfloor \frac{k}{T^{(l)}} \bigr \rfloor$. Furthermore, if the landmark is detectable over any horizon, then the landmark estimate is also RGES.
\end{remark}

\section{Discussion} 
\label{sec:4}

In this section, we introduce two specific sensor configurations, discuss their compatibility with the MHE SLAM scheme from Section~\ref{sec:3}, and propose a simplified scheme applicable when range measurements of landmarks are available.

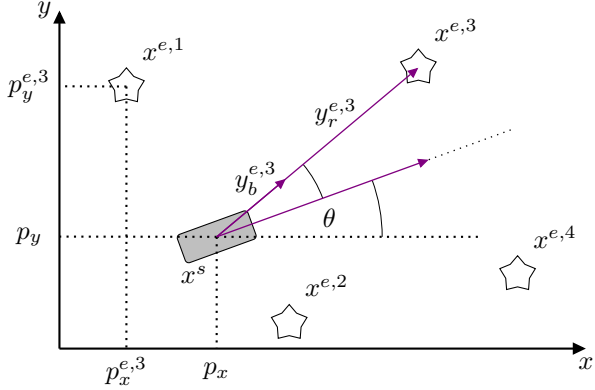
\begin {figure}%[!hbtp]
\vspace*{0.1cm}
\centering
\begin{adjustbox}{max height=0.55\textwidth, max width=0.48\textwidth}
\begin{tikzpicture}[scale=1.0]

    \coordinate (P1) at ($(0,0) + (20:1.5cm)$);
    \coordinate (P2) at ($(0,0) + (200:1.5cm)$);
    \coordinate (Theta) at ($(P2) + (1.5, 0.25)$);
    \coordinate (L1) at ($(P2) + (-1.2, 2)$);
    \coordinate (L2) at ($(P2) + (310:1.5cm)$);
    \coordinate (L3) at ($(P2) + (40:3.5cm)$);
    \coordinate (L4) at ($(P2) + (4, -0.5)$);

    \coordinate (L1l) at ($(L1) + (+0.5, +0.5)$);
    \coordinate (L2l) at ($(L2) + (+0.5, +0.5)$);
    \coordinate (L3l) at ($(L3) + (+0.5, +0.5)$);
    \coordinate (L4l) at ($(L4) + (+0.5, +0.5)$);

    \coordinate (Ls) at ($(P2) + (-0.3, -0.5)$);

    \coordinate (M3) at ($(P2) + (40:1.2cm)$);
    \coordinate (M3l) at ($(M3) + (-0.4, 0)$);
    \coordinate (M3ll) at ($(L3) + (-1.1, -0.6)$);

    \node[draw, rectangle, rounded corners=0.5mm, minimum width=1cm, minimum height=0.4cm, anchor=center, fill=lightgray, rotate=20] at (P2) {};
    \draw[line width=0.8pt, dotted] (P2) -- (-1.41, -2);
    \draw[line width=0.8pt, dotted] (P2) -- (-3.5, -0.51);
    \draw[line width=0.8pt, dotted] (P2) -- ++(3.5, 0);

    \draw[line width=0.8pt, dotted] (L1) -- (-2.61, -2);
    \draw[line width=0.8pt, dotted] (L1) -- (-3.5, 1.49);

    %%%%% Angle lines %%%%%

    \draw[line width=0.5pt, dotted] (P1) -- ++(20:1.2cm);
    \draw[line width=0.3pt] (P2) ++(0:2.2cm) arc[start angle=0, end angle=20, radius=2.2cm];
    \draw[line width=0.3pt] (P2) ++(20:1.5cm) arc[start angle=20, end angle=40, radius=1.5cm];
    
    \draw[-{Triangle[length=4pt, width=3.5pt]},line width=0.5pt, violet] (P2) -- ++(20:3cm);
    \node[ scale=1]() at (Theta)   {$\theta$};
    \node[scale=1, draw, star] at (L1){};
    \node[scale=1, draw, star] at (L2){};
    \node[scale=1, draw, star] at (L3){};
    \node[scale=1, draw, star] at (L4){};

\coordinate (L2) at ($(P2) + (310:1.5cm)$);
    \draw[-{Triangle[length=4pt, width=3.5pt]},line width=0.5pt, violet] (P2) -- (L3);
    \draw[-{Triangle[length=4pt, width=3.5pt]},line width=0.5pt, violet] (P2) -- (M3);
    
    %%%%% Coord syst %%%%%

    \draw[-{Triangle[length=5pt, width=4.5pt]},line width=0.8pt](-3.5, -2)--(3.5,-2);
    \draw[-{Triangle[length=5pt, width=4.5pt]},line width=0.8pt](-3.5, -2)--(-3.5, 2.5);
    
    \node[ scale=1]() at (3.5,-2.2)  {$x$};
    \node[ scale=1]() at (-3.7, 2.5)   {$y$};
    \node[ scale=1]() at (-1.41, -2.3)   {$p_x$};
    \node[ scale=1]() at (-3.9, -0.51)   {$p_y$};
       \node[ scale=1]() at (-2.61, -2.3)   {$p_x^{e,3}$};
    \node[ scale=1]() at (-3.9, 1.49)   {$p_y^{e,3}$};

    \node[ scale=1]() at (Ls)  {$x^{s}$};

    \node[ scale=1]() at (M3l)  {$y^{e,3}_b$};
    \node[ scale=1]() at (M3ll)  {$y^{e,3}_r$};
    
    \node[ scale=1]() at (L1l)  {$x^{e,1}$};
    \node[ scale=1]() at (L2l)   {$x^{e,2}$};
    \node[ scale=1]() at (L3l)   {$x^{e,3}$};
    \node[ scale=1]() at (L4l)   {$x^{e,4}$};

    % \fill (0,0) circle (1pt);
    
\end{tikzpicture}
\end{adjustbox}
    \caption{Planar robot and four landmarks. Coordinates of the robot are denoted by $x^s=(p_x^{\top}, p_y^{\top}, \theta^{\top})^{\top}$, and the coordinates of each landmark $l$ with $x^{e,l}=(p_x^{e,l\top}, p_y^{e,l \top})^{\top}$. Measurement of landmark $l=3$ coming from a bearing-only sensor is a unit vector denoted by $y_b^{e,3}$, while the measurement obtained with a range sensor is denoted as $y_r^{e,3}$. }
\label{fig:img4}
\end{figure}

\subsection{Common SLAM Configurations}

Different types of SLAM setups exist depending on the available sensors. For example, Visual-Inertial SLAM \cite{campos2021orb} relies on visual and inertial data, while lidar-based SLAM might use only range measurements \cite{montemerlo2003fastslam} or combine them with inertial data \cite{hess2016real}. In the case of camera-based SLAM, the type of camera introduces further distinctions: monocular cameras provide  bearing-only measurements, while stereo cameras provide range measurements, similar to lidar. These two types of measurements, i.e., bearing-only or range, represent the most frequently encountered landmark measurement models, despite various possible sensor combinations. In the following, we will briefly introduce these models, as well as an example of a SLAM system as introduced in \eqref{eq:slam_sys}.

We consider a model of a planar robot from \cite{flayac2023distributed}, shown in Figure \ref{fig:img4} and described by   %$x^s(k+1) = f(x^s(k), u(k), v(k))$
\begin{equation}
    \begin{aligned}
        p_{x,{k+1}} &= p_{x,{k}} + u_{1,k}\cos{\theta_k} + v_{1,k},\\
        p_{y,{k+1}} &= p_{y,{k}} + u_{1,k}\sin{\theta_k} + v_{2,k},\\
        \theta_{k+1} &= \theta_k + u_{2,k} + v_{\theta,k},
    \end{aligned}
    \label{eq:kin_dyn}
\end{equation}
where $x^s_k=(p_{x,k}^{\top}, p_{y,k}^{\top}, \theta_k^{\top})^{\top}$, $u_k=(u^{\top}_{1,k}, u^{\top}_{2,k})^{\top}$ and $v_k=(v_{1,k}^{\top}, v_{2,k}^{\top}, v_{\theta,k}^{\top})^{\top}$. The states $p_{x,k}$ and $ p_{y,k}$ correspond to $x$ and $y$ coordinates of the robot in the plane at time $k$, $ \theta_k$ represents its orientation, while $u_{1,k}$ and $u_{2,k}$ are linear and angular velocities at time $k$.

Measurements of ego-state are given by 
\begin{equation}
    y^s_{k} = x^s_{k} + \xi^s_k,
    \label{eq:sys_measurement}
\end{equation}
and as will be discussed later in this section, measurements analogous to these are necessary to ensure observability of the SLAM problem. While they may be obtained by external sensors, they may also be a result of e.g., sensing that relies on landmarks with known position.

For two popular landmark measurement models that encompass the most popular types of SLAM, i.e., monocular, stereo and lidar-based, the measurement $y_k^{e,l}$ of landmark $l$ is given by
\begin{equation}
y_k^{e,l}=a_k^{e,l}\left(R\left(-\theta_{k}\right) \frac{x^{e,l}-p_k}{\alpha(x^{e,l},p_k)}+\xi_k^{e,l}\right).
\label{eq:bearing_only}
\end{equation}
The choice of $\alpha$ depends on the measurement model, $R(\theta)$ is a rotation matrix for angle $\theta$ and $p_k\coloneqq(p_{x,{k}}^{\top}, p_{y,{k}}^{\top})^{\top}$.

For the bearing-only measurement, it holds that $\alpha(x^{e,l},p_k) = {\left\|x^{e,l}-p_k\right\|}$. This measurement model corresponds to a monocular camera, because due to the lack of depth information from the image, it only provides a ray along which the landmark is located. For the range measurement, $\alpha(x^{e,l},p_k) = 1$. This measurement model corresponds to a lidar or a stereo camera and is capable of providing the distance to the measured landmark along a ray.

\subsection{Verification of Assumptions}

 A systematic method to compute an i-IOSS Lyapunov function and thus to verify Assumption \ref{ass1}  is proposed in \cite{schiller2023lyapunov}, where Remark 1 comments on the choice of matrices introduced in Definition \ref{def1}.  Since observability implies detectability \cite{muntwiler2025nonlinear}, the satisfaction of Assumption \ref{ass1} is implied whenever the problem is observable. In the context of SLAM, observability has been studied extensively for the linearized systems used in EKF updates \cite{andrade2004effects, bayat2012observability, le2017observers}, as well as from the perspective of nonlinear observability \cite{lee2006observability}. From \cite{andrade2004effects}, we have that, without ``anchoring'', or an external sensor measuring the position of the robot, even a linear system with linear landmark measurements is not observable. Furthermore, in \cite{lee2006observability} it was shown that the standard 2D planar SLAM formulation relying on range measurements of landmarks is inherently unobservable, and that the observability can be achieved by ensuring the measurements of at least two landmarks with known positions are available (similar to ``anchoring'' from \cite{andrade2004effects}). In this case, one would consider the measurements corresponding to these anchor points as part of the state measurement \eqref{eq:slam_sys2}, and  Assumption \ref{ass1} would be satisfied. Therefore, one can rely on the results from the literature to ensure observability and hence satisfaction of Assumption \ref{ass1}. Furthermore, if any set of landmarks is always persistently excited, the system state can be augmented with the positions of these landmarks. Specifically, for our example, due to linear system dynamics and linear measurement model, it can be directly shown by checking the rank of the observability matrix that the system is observable and therefore also detectable.

Assuming the existence of the minimal and maximal sensor range, where both are strictly positive, Assumption \ref{ass2} is satisfied for both range and bearing-only measurement models.

Sufficient conditions for satisfaction of condition \eqref{eq:ass3} can also be derived directly from observability results. For example, if it is known that a landmark is observable when always visible over a horizon of certain length, then  condition \eqref{eq:ass3} is satisfied for every horizon of the same length over which the visibility of the landmark is maintained. However, using such sufficient conditions might result in overly conservative update schemes. Therefore, a  better understanding of necessary and sufficient conditions for landmark detectability in SLAM is an interesting future research problem.

\subsection{Streamlined Method for the Range Measurement Model}
\label{sec:5}

In the case of the range measurement model, the landmark measurements \eqref{eq:bearing_only} are linear in landmark coordinates. As a result, the landmark position update from \eqref{eq:Pkle} becomes an unconstrained, linear optimization problem that can be solved using a recursive least squares approach whenever $a_k^{e,l}=I_d$. This aligns with the setup in \cite{bin2022generalized}, where $\phi^*(t) = R(-\hat{\theta_t})$, $ d_{\phi}(t)=0$ and $d_y(t)=\overline{\xi}_{t}^{e,l}$ from \eqref{eq:rewrite_meas}, and therefore yields the same stability properties (i.e., RGES) of the landmark estimation error with respect to $\overline{\xi}_{j}^{e,l}$ and modified time index that corresponds to the number of time steps when the landmark is visible. Consequently, since $\overline{\xi}_{j}^{e,l}$ is upper-bounded by the state estimation error and measurement noise, the recursive least squares approach results in a bound similar to the one in \eqref{eq:prop1}. In this case, verifying landmark detectability reduces to confirming the existence of an upper and lower bound for $\phi^*(t)\phi^*(t)^{\top}$, which is always satisfied when a landmark is visible due to the properties of a rotation matrix.  As a result, the same stability properties are achieved as with an MHE scheme, while the computational load is significantly reduced.

\section{Simulation Results}
\label{sec:6}

\begin{figure}[h]   % Optional [h] to place the figure here
        \centering
        \includegraphics[width=\linewidth]{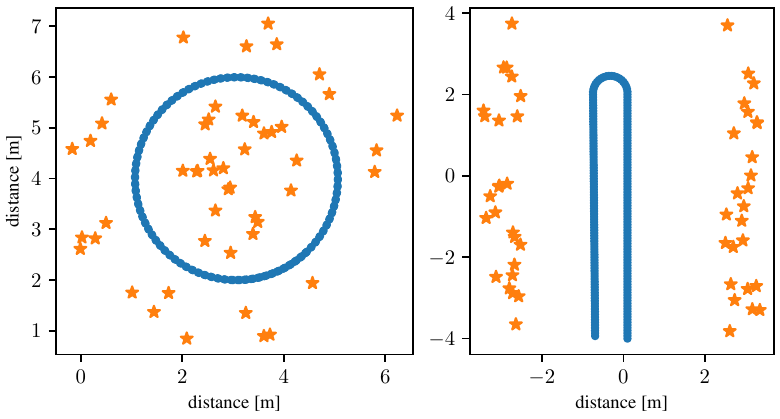} 
    \caption{ Circular (left) and corridor (right) landmark placement (indicated by stars) and corresponding system trajectories.}
    \label{fig:landmarks}
\end{figure}

In this section, we present simulations of a planar robot with dynamics described in \eqref{eq:kin_dyn}, state measurements as in \eqref{eq:sys_measurement} and bearing-only measurement given by \eqref{eq:bearing_only} in two different environments similar to those introduced in \cite{flayac2023distributed}. Both environments are featured by $L=50$ landmarks, arranged in a circular or corridor-like manner, while the robot moves on a circular or straight trajectory, as shown in Figure \ref{fig:landmarks}. 

We simulate measurement and process noise with standard deviation of $0.01$, initialize all landmark coordinates to $(0,0)$ and set the initial value of the robot ego-state to its true value. Furthermore, we set $\overline{U}^s=0.001I_3, Q^{s}=I_6, R^{s}=I_3, \overline{U}^{e}=0.01I_2, Q^{e,l}=I_2, R^{e,l}=0.1I_2$, where  $l \in \mathbb{I}_{[1,L]}$. We compare our method to the coupled update based on MHE for the augmented state described in Section \ref{sec:2}, where all corresponding weight matrices are the same. In all MHE schemes, $\eta=0.99$ and horizon length is $20$. Furthermore, the range of the sensor is set to $2$ m for the circular scenario and $3$ m for the corridor scenario.

\begin{figure}[h]   % Optional [h] to place the figure here
        \centering
        \vspace*{0.2cm}
        \includegraphics[width=\linewidth]{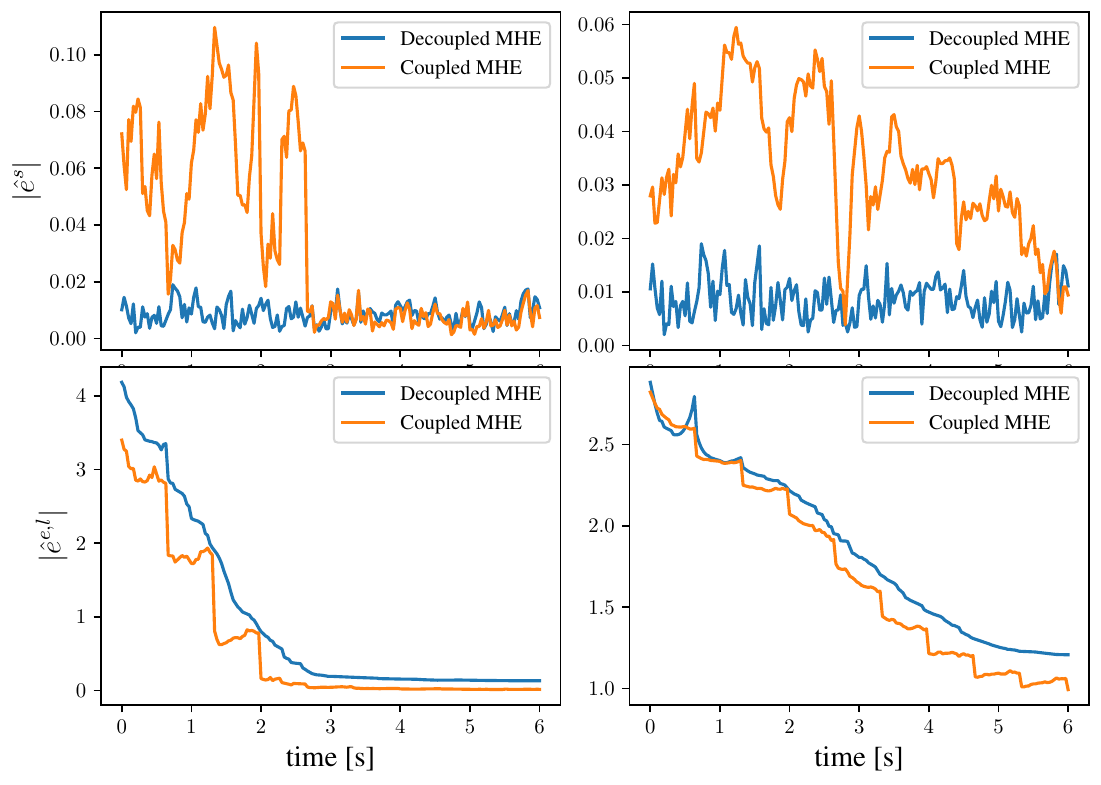}  % Replace with your image path
    \caption{Ego-state (top) and average landmark error (bottom) for the circular (left) and corridor (right) scenarios. }
    \vspace*{-0.5cm}
    \label{fig:straight}
\end{figure}

The resulting average landmark and ego-state estimation errors are shown in Figure \ref{fig:straight}. The decoupled approach results in better ego-state estimates compared to the coupled approach in both scenarios, while landmark estimates remain comparable.  The effect is stronger in the corridor scenario, as the lack of parameter excitation is more frequent, causing the coupled MHE  to use insufficient information for state and landmark estimates. Furthermore, the proposed decoupled approach results in significantly lower computation times: in the corridor scenario, the coupled scheme update takes on average 8.5 s to solve, while the decoupled scheme takes 9.8 ms for the ego-state update and 0.4 ms per landmark update. This highlights that a decoupled approach is much better suited for real-time applications, while still capable of providing bounded error guarantees whenever the assumptions are satisfied. 

\section{Conclusion}
\label{sec:7}
This paper introduced an MHE scheme for SLAM, analyzed
its robustness, and verified the necessary assumptions
in the context of realistic SLAM setups featuring specific
measurement models and presence of noise. The scheme addresses loss of detectability, while at the same time also allowing for parallelization of landmark updates. This is achieved by decoupling the ego-state and landmark  estimates, where the latter are  performed only if the landmark is detectable over the most recent horizon. It is shown that ego-state estimates are RGES, while the error bounds on landmark estimates depend on how often each landmark is detectable. Future work includes expanding the conducted analysis further to establish stability properties of the coupled estimation scheme and using the derived error bounds in an active SLAM setting.

\bibliographystyle{IEEEtran}
\bibliography{references.bib}

\end{document}